\documentclass[aps,prl,reprint,twocolumn,floatfix,showpacs,superscriptaddress]{revtex4-1}
\pdfoutput=1

\usepackage[author={pi},hspace=20px,icon=Comment,color=yellow]{pdfcomment}
\usepackage{graphicx}
\usepackage{amsmath,amssymb}
\usepackage{natbib}
\usepackage{hyperref}
\usepackage{xcolor}

\usepackage[T1]{fontenc} 

\begin{document}


\title{Multimode plasmon excitation and \emph{in-situ} analysis in top-down fabricated nanocircuits}

\author{Peter Geisler}
\thanks{equally contributing}
\author{Gary Razinskas}
\thanks{equally contributing}
\author{Enno Krauss}
\author{Xiao-Fei Wu}
\affiliation{NanoOptics \& Biophotonics Group, Experimentelle Physik 5, Physikalisches Institut, Universit\"at W\"urzburg, Am Hubland, 97074 W\"urzburg, Germany}
\author{Christian Rewitz}
\author{Philip Tuchscherer}
\author{Sebastian Goetz}
\affiliation{Institut f\"ur Physikalische und Theoretische Chemie,\\Universit\"at W\"urzburg, Am Hubland, 97074 W\"urzburg, Germany}
\author{Chen-Bin Huang}
\affiliation{Institute of Photonics Technologies, National Tsing Hua University, Hsinchu 30013, Taiwan}
\author{Tobias Brixner}
\email[]{brixner@phys-chemie.uni-wuerzburg.de}
\affiliation{Institut f\"ur Physikalische und Theoretische Chemie,\\Universit\"at W\"urzburg, Am Hubland, 97074 W\"urzburg, Germany}
\affiliation{R\"ontgen Research Center for Complex Material Systems (RCCM), Am Hubland, 97074 W\"urzburg, Germany}
\author{Bert Hecht}
\email[]{hecht@physik.uni-wuerzburg.de}
\affiliation{NanoOptics \& Biophotonics Group, Experimentelle Physik 5, Physikalisches Institut, Universit\"at W\"urzburg, Am Hubland, 97074 W\"urzburg, Germany}
\affiliation{R\"ontgen Research Center for Complex Material Systems (RCCM), Am Hubland, 97074 W\"urzburg, Germany}

\date{\today}

\begin{abstract}
We experimentally demonstrate synthesis and \emph{in-situ} analysis of multimode
plasmonic excitations in two-wire transmission lines supporting a symmetric and
an antisymmetric eigenmode. To this end we irradiate an incoupling antenna with
a diffraction-limited excitation spot exploiting a polarization- and
position-dependent excitation efficiency. Modal analysis is performed by
recording the far-field emission of two mode-specific spatially separated
emission spots at the far end of the transmission line. To illustrate the power
of the approach we selectively determine the group velocities of symmetric and
antisymmetric contributions of a multimode ultrafast plasmon pulse.
\end{abstract}


\maketitle
Miniaturization of functional optical circuits is hampered by the
diffraction-limited modal profiles of dielectric waveguides
\cite{yariv_photonics_2007}. In contrast, plasmonic modes supported by
noble-metal nanostructures offer subwavelength confinement
\cite{novotny_light_1994, takahara_guiding_1997, barnes_surface_2003} and
therefore promise the realization of nanometer-scale integrated optical
circuitry with well-defined functionality \cite{ozbay_plasmonics:_2006,
gramotnev_plasmonics_2010}. While single-mode operation is a design goal for
dielectric waveguides, in plasmonic nanocircuits multimode interference could
lead to enhanced functionality based on the control of near-field intensity
patterns \cite{stockman_coherent_2002, sukharev_phase_2006,
tuchscherer_analytic_2009}.

Previous work towards the realization of optical nanocircuits relied on
chemically grown single-wire waveguides pioneered by Ditlbacher \emph{et al.}
\cite{ditlbacher_silver_2005}. In such systems the plasmon excitation and
emission efficiencies depend on the wire diameter, which at the same time also
determines the spectrum of modes and their respective dispersion relations. In
addition, structural uncertainties, such as the uncontrolled shape of the end
facets, have a strong influence on the far-field excitation and emission
properties of different modes \cite{li_correlation_2010,
shegai_unidirectional_2011, song_imaging_2011}.  By combining careful selection
and micromanipulation of chemically grown nanowires, indeed prototypes of
optical nanocircuitry have been demonstrated in which multimode interference is
exploited \cite{fang_branched_2010, wei_cascaded_2011}.

Despite these achievements, it is a necessary next step towards advanced
plasmonic nanodevices to obtain independent control over light coupling and
propagation in optical nanocircuits by a deterministic synthesis of multimodal
excitations. The existence of a transverse and a longitudinal mode in chains of
closely spaced plasmonic nanoparticles \cite{quinten_electromagnetic_1998,
maier_local_2003} led to a proposal of deterministic coherent control of a
routing functionality in a branched particle-chain waveguide circuit
\cite{tuchscherer_analytic_2009}. Later on, easier to fabricate
metal-insulator-metal (MIM)-type waveguides, such as grooves and channels
\cite{bozhevolnyi_channel_2005} were applied e.g., to implement logic
operations, albeit via single-mode interference \cite{fu_all-optical_2012}. 

To obtain more flexibility and control in terms of excitation schemes and
available modes, two-wire transmission lines (TWTLs) offering a symmetric and an
antisymmetric mode \cite{ly-gagnon_routing_2012}, are a logical extension of
both single-wire and MIM concepts. However, experiments so far have only considered 
the antisymmetric mode\cite{krenz_near-field_2010,schnell_nanofocusing_2011,ly-gagnon_routing_2012}. 
Importantly, TWTLs can be combined with linear dipole antennas 
to tailor the in- and outcoupling of light \cite{huang_impedance_2009, wen_excitation_2009, krenz_near-field_2010,
schnell_nanofocusing_2011}.

Here, we use plasmonic nanocircuits \cite{huang_impedance_2009} consisting of an optimized incoupling antenna 
(generator), a transmission line with a mode-dependent characteristic impedance, and a 
mode detector (load). The structures are fabricated by focused ion beam (FIB) milling of
single-crystalline gold flakes and therefore structural detail and plasmon propagation are not limited by the 
grain boundaries of multi-crystalline gold films \cite{huang_atomically_2010,ditlbacher_silver_2005}.
We show that by controlling polarization and exact position of a near-infrared laser focus
with respect to the antenna, any superposition of TWTL modes can be launched
and analyzed by a single-shot \emph{in-situ} far-field read-out of the mode
detector. As a first application, we selectively determine group velocities and
time delays of pure-mode contributions of a multimode ultrafast TWTL plasmon
pulse.

Control of multimodal excitations provides advanced circuit functionality, e.g.
through deterministic coherent control of nanooptical fields as described in
\cite{tuchscherer_analytic_2009} where multimode operation is a necessity. 
Such control of confined fields can also be used for the implementation of
nonlinear optical switching effects and the controlled interaction of guided
modes with single quantum emitters \cite{akimov_generation_2007,kolesov_waveparticle_2009}.

\begin{figure}[tbp]
\begin{center}
\includegraphics[width=0.8\columnwidth]{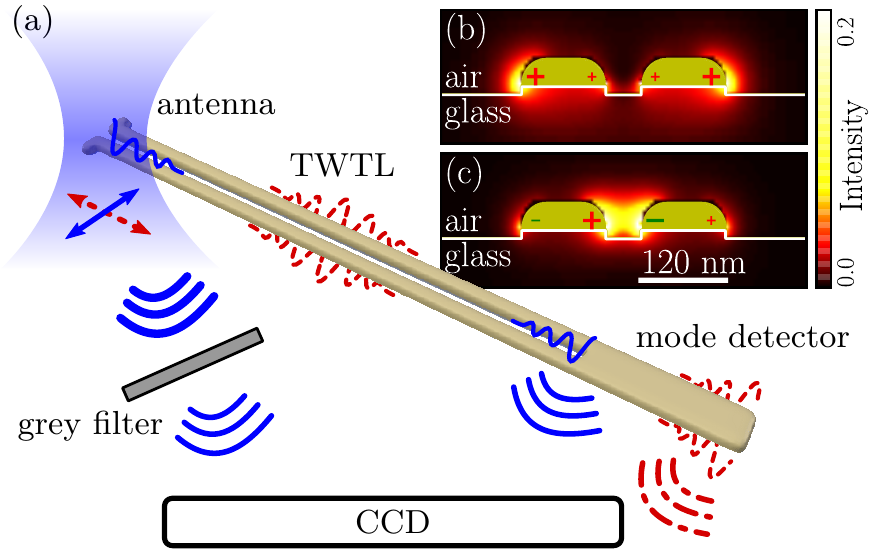}
\caption{(color online) Principle of the experiment. (a) Cartoon: launching,
	propagation and re-emission of TWTL modes. (b,c) Simulated transverse modal
	profiles in an infinitely long TWTL on glass for the symmetric (b) and
	antisymmetric eigenmode (c). Instantaneous charge distributions are
	symbolized by ``+'' and ``-''. 
\label{fig:modeprofiles}}
\end{center}
\end{figure}

Plasmonic nanocircuits [Fig. \ref{fig:modeprofiles} (a)] consisting of an
incoupling antenna, a TWTL, and a mode detector, were fabricated by FIB milling
(FEI company, Helios NanoLab) of a large single-crystalline gold flake (38~nm
thickness) deposited on a cover glass \cite{huang_atomically_2010}. For the
chosen dimensions only two TWTL modes with considerable propagation length
(>2~$\mu$m) exist. The transverse modal profile of any multimodal excitation
propagating along the TWTL at a fixed position therefore is a superposition of
these two eigenmodes after transients have expired \cite{huang_impedance_2009}.
\begin{table}
	\begin{ruledtabular}
		\begin{tabular}{l r r}
			mode property & symmetric & antisymmetric\\
			$\lambda_{\text{eff}} / \mathrm{nm}$& 480& 434\\ 
			$l_{\text{decay}} / \mathrm{nm}$& 2628& 1892\\ 
			$v_g / 10^8 \mathrm{\,m\,s^{-1}}$& 1.26& 1.16\\ 
		\end{tabular}
	\end{ruledtabular}
	\caption{Eigenmode parameters (free-space wavelength 830~nm).
		$\lambda_{\text{eff}}$ denotes the effective wavelength,
		$l_{\text{decay}}$ the intensity decay length and $v_g$ the group
		velocity of either mode, respectively.
	\label{tab:mode-properties}}
\end{table}
Figures \ref{fig:modeprofiles} (b,c) show the simulated transverse eigenmode
profiles of the symmetric and the antisymmetric mode for a TWTL consisting of a
pair of gold wires with rounded upper corners on glass elevations above a glass
half space (Lumerical Solutions Inc., MODE Solutions). The false color scale
represents the time-averaged near-field intensity. The corresponding eigenmode
properties are summarized in Table \ref{tab:mode-properties}. The longer
propagation length, longer effective wavelength, as well as higher group
velocity observed for the symmetric mode are consistent with its lower field
confinement.

The mode detector attached to the far end of the TWTL can be read out via a
diffraction-limited far-field measurement and therefore provides direct
\emph{in-situ} feedback on the actual modal composition. Its operation principle
relies on the different interaction of symmetric and antisymmetric modes with
different types of TWTL discontinuities. Due to a field node in the gap the
symmetric eigenmode is not affected by a termination of the gap while it is
strongly reflected and radiated at a complete termination of the circuit. The
antisymmetric eigenmode exhibits a field maximum in the gap and is thus strongly
reflected and radiated as soon as the gap is terminated. A sequence of a gap
shortcut followed by a complete termination of the TWTL therefore acts as a mode
detector by spatially separating the respective emission spots.

\begin{figure}[htbp]
\begin{center}
\includegraphics[width=\columnwidth]{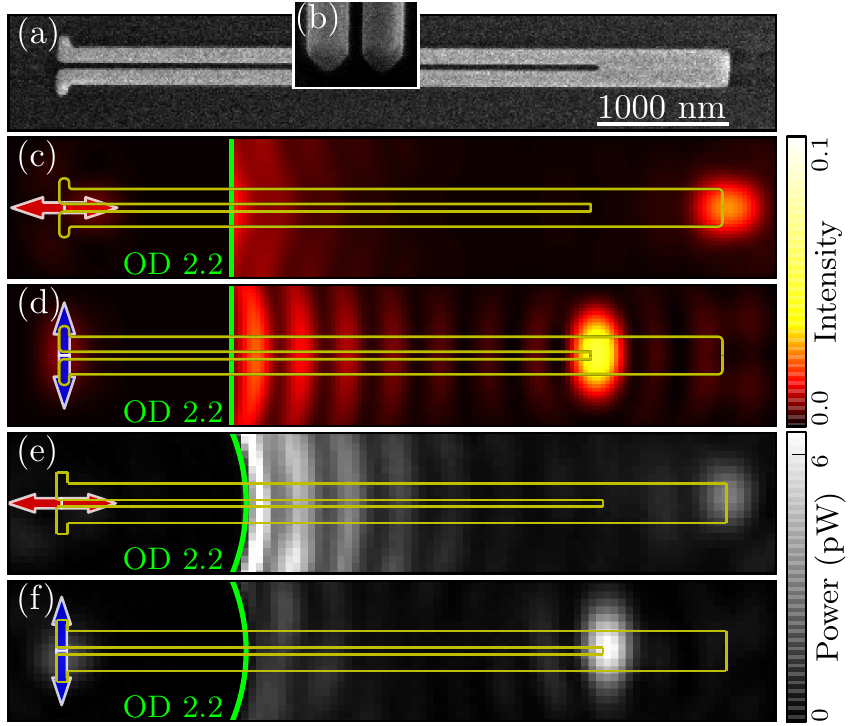}
\caption{(color online) (a) SEM of the plasmonic nanocircuit. (b) SEM cross
	section of a TWTL (52$^\circ$ observation angle). 
(c,d) Simulated far-field images (far-field projection
\cite{lumerical_inc._imaging_2012}) for a pure symmetric (c) and antisymmetric
mode (d) arriving at the mode detector. Pure modes are excited by illuminating
the incoupling antenna (overall length 450~nm) with a focused beam polarized
parallel (c) and perpendicular (d) to the TWTL. The intensity scale is
normalized to the reflected intensity at the glass/air interface. To match the
experiments the antenna reflections have been attenuated numerically using
OD~2.2 (left of the green line).
(e,f) CCD images (detected power per pixel) showing the attenuated (OD~2.2,
indicated by the green circle) reflected spot at the incoupling antenna position
and the mode detector emission spots (100~nW excitation power). (c-f) The arrows
indicate the polarization. A nanocircuit outline is superimposed as a guide to
the eye. All images show background ``fringes'' from direct laser scattering
which have no influence on the transmission and thus the outcome of the
experiment.
\label{fig:transmission}}
\end{center}
\end{figure}

Figure \ref{fig:transmission} (a,b) show scanning electron micrographs (SEM) of
a plasmonic nanocircuit featuring a 4~$\mu$m long TWTL visualizing the rounded
shape of the TWTL wires caused by secondary sputtering processes during FIB
milling. Figures \ref{fig:transmission} (c,d) show far-field FDTD simulations of
the mode detector interacting with either of the two eigenmodes.

The 1~$\mu$m spacing between the discontinuities results in two clearly
separated emission spots whose intensities are proportional to the amplitudes of
the respective eigenmode contributions. The emission is polarized parallel to
the wire axis for the symmetric [Fig. \ref{fig:transmission} (c)] and
perpendicular for the antisymmetric mode emission spot [Fig.
\ref{fig:transmission} (d)]. For a quantitative determination of the power in
each mode the respective radiation efficiencies of the two emission points have
to be taken into account. As a figure of merit (FOM) for the ability of the mode
detector to separate modal contributions we define for an incoming pure
symmetric (antisymmetric) mode plasmon ${\rm FOM}_{\rm i} = I_{\rm i}/\sum_{\rm
i}I_{\rm i}\; , \;{\rm i} = \{\text{sym.}, \text{antisym.}\}$, 
where $I_{\rm i}$ is the emission intensity at the symmetric (antisymmetric)
emission spot. Perfect mode selectivity corresponds to ${\rm FOM}_{\rm i} = 1$.
Due to the small amount of scattered light at the position of the mode detector
we achieve typical figure of merits of 0.98 in experiments and in simulations. 

In order to experimentally characterize the launching and emission of plasmon
excitations, the cover glass supporting the nanocircuit is mounted onto a
home-built inverted microscope setup. An oil immersion microscope objective
(Leica, 1.30 NA, $\infty$, PL Fluotar 100x) is used to focus a linearly
polarized laser beam ($\lambda$ = 830~nm, 12~nm FWHM spectral linewidth, 80~MHz
repetition rate, 100~nW average power measured in front of the objectives back
aperture, NKT Photonics, SuperK Power with SpectraK AOTF) via a
$\lambda/2$-plate (Foctec, AWP210H NIR) to a diffraction-limited (390~nm
diameter) spot at the glass/air interface. Once the spot is overlapped with the
incoupling antenna of a TWTL structure, plasmons are excited and subsequently
re-emitted at the far end of the structure. The same objective is used to image
the emission spots onto a CCD camera (Andor, DV887AC-FI EMCCD) via a 50/50
nonpolarizing beamsplitter (Thorlabs, CM1-BS013). In order to avoid saturation
of the CCD, the strong reflection of the excitation spot is suppressed by a
small beam block (OD 2.2) introduced in an intermediate image plane. The exact
position of the excitation spot can be adjusted with nm-precision by moving the
sample using a piezo translation stage (Physik Instrumente, P-527). The
principle of the experiment is sketched in Fig. \ref{fig:modeprofiles} (a).
Figures \ref{fig:transmission} (e) and (f) show far-field images of the
structure being excited at the antenna and re-emitting light at the mode
detector structure for excitation polarizations parallel (e) and perpendicular
(f) to the TWTL. To launch the antisymmetric mode in (f) the polarization was
rotated by 90$^\circ$ while the excitation spot was kept fixed. This
demonstrates the possibility to excite a well-defined superposition of both
modes simply by adjusting the laser polarization. It is interesting to observe
that not only the positions of the far-field emission spots match very well the
simulations in Figs. \ref{fig:transmission} (c,d) but also the respective spot
shapes.

\begin{figure}[htbp]
\begin{center}
\includegraphics[width=\columnwidth]{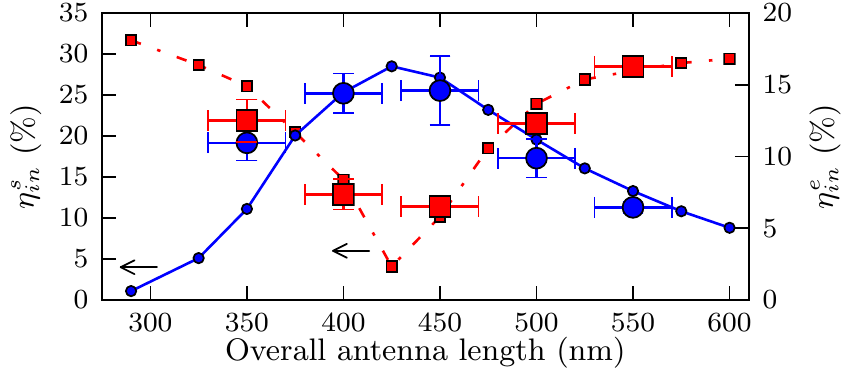}
\caption{(color online) Simulated (lines, small symbols, $\eta_{\text{in}}^{s}$)
and experimentally obtained incoupling efficiency (large symbols with error
bars, $\eta_{\text{in}}^{e}$) vs. total antenna length. Symmetric mode:
Illumination parallel to the TWTL (red squares, dashed line). Antisymmetric
mode: Illumination perpendicular to the TWTL (blue circles, solid line). 
\label{fig:incoupling}}
\end{center}
\end{figure}

The excitation efficiency of both TWTL modes can be engineered by utilizing an
incoupling structure that links the field profile of the excitation beam to the
modal profiles of the TWTL. Here we use a simple dipole antenna where the length
and width of the antenna arms influence the antenna impedance and therefore the
transfer of power to the respective waveguide mode \cite{huang_impedance_2009}.
In Fig. \ref{fig:incoupling} we plot both the experimentally determined and
simulated incoupling efficiencies into either TWTL mode as a function of the
overall antenna length (350 - 550~nm, the 290~nm simulation corresponds to the
``no-antenna'' -- ``wire-only'' case) for a fixed antenna width (80~nm). In
simulations, the mode-specific incoupling efficiency $\eta_{\text{in}}^{s}$ is
obtained by evaluating a mode overlap integral at a distance $x$ from the
antenna where direct influence of the excitation beam and transients can be
neglected (i.e., 2.5~$\mu$m from the antenna). By taking into account the
simulated decay length (see Table \ref{tab:mode-properties}) this power is then
extrapolated towards the incoupling antenna position. Consequently,
$\eta_{\text{in}}^{s}$ reads as
\begin{equation*}
	\eta_{\text{in}}^{s} = \frac{p(x)}{p_0} \times \left[e^{- x / l_{\text{decay}}}\right]^{-1} \; ,
\end{equation*}
where $p(x)$ is the power in the respective mode at position $x$ along the wire,
$p_0$ the laser power and $l_{\text{decay}}$ the intensity decay length of the
mode. For polarization parallel to the TWTL (Fig. \ref{fig:incoupling}, red
dashed line) the incoupling efficiency of the symmetric mode decreases from a
large value of >30\% for the case without antenna to a narrow minimum of <5\% at
around 425~nm overall antenna length and then largely recovers for a further
increase of the antenna length. Excitation of the antisymmetric mode is symmetry
forbidden for this polarization unless the focus is displaced (see below). For a
polarization perpendicular to the TWTL (Fig. \ref{fig:incoupling}, blue solid
line) the behavior is quite the opposite. Without antenna the antisymmetric mode
can hardly be excited (<2\% excitation efficiency). By increasing the antenna
length to about 425~nm the incoupling efficiency reaches a maximum of almost
30\% and then decreases towards another minimum for even longer antennas. For
perpendicular polarization, coupling to the symmetric mode is symmetry
forbidden.

To experimentally verify these predictions, 2 nominally identical arrays
consisting of 5 nanocircuits with scanned antenna lengths have been fabricated.
By taking into account the damping of the wire (see Table
\ref{tab:mode-properties}) and the radiation efficiencies of the mode detector
emission spots, the power in each mode at the antenna position can be
extrapolated. We plot the experimental incoupling efficiency $\eta_{in}^{e}$
(Fig. \ref{fig:incoupling}, large symbols), defined as
\begin{equation*}
	\eta_{\text{in}}^{e} = \frac{p_{\text{out}}}{p_0} \times \left[\eta_{\text{out}} \times e^{- L / l_{\text{decay}}}\right]^{-1} \; ,
\end{equation*}
where $p_{\text{out}}$ is the integrated emitted power at an emission spot,
$p_0$ the excitation power measured in front of the objective,
$\eta_{\text{out}}$ the radiation efficiency of either emission spot corrected
by the collection solid angle of the objective lens (about 25\% for both modes),
$L$ the length of the TWTL, and $l_{\text{decay}}$ the simulated decay length.
While the experimental data very well reproduce the general trend the absolute
experimental values for the incoupling efficiency are smaller by about a factor
of 2. Such a deviation can be caused by experimental decay lengths that are
about 25\% shorter than predicted by simulations. The reason for such deviations
is unclear but effects like surface scattering of electrons likely contribute.

\begin{figure}[htbp]
\begin{center}
\includegraphics[width=\columnwidth]{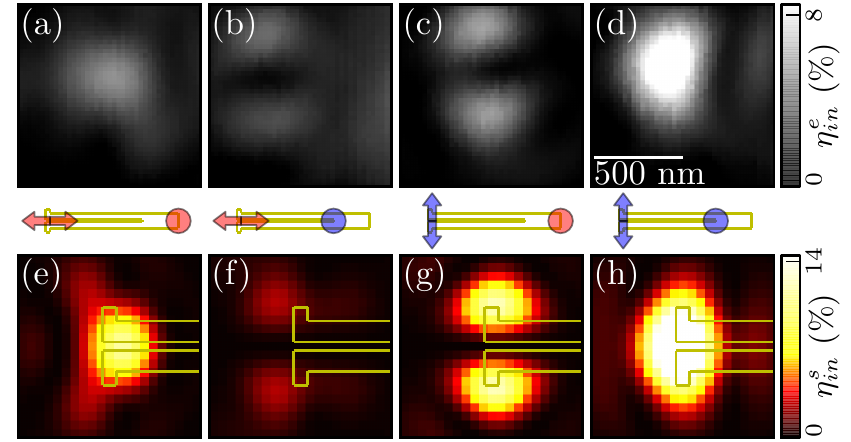}%
\caption{(color online) Focal spot position dependent modal excitation
	efficiency maps (450~nm overall antenna length): (a--d) experiments, (e--h)
	simulations. Middle panel: pictograms indicating excitation polarization and
	emission position. When centered on the antenna the antisymmetric mode (d,h)
	is more efficiently excited than the symmetric mode (a,e) consistent with
	Fig. \ref{fig:incoupling}.
\label{fig:position}}
\end{center}
\end{figure}

So far we have assumed an excitation focal spot perfectly centered on the
incoupling antenna (neglecting small displacements along the TWTL) leading to
the excitation of pure modes for the two fundamental polarizations. We test the
stability of such a configuration by recording the excitation efficiencies into
both modes for parallel and perpendicular polarization as a function of beam
displacements.  Breaking the symmetry,  such displacements lead to a significant
increase in the excitation efficiency of the respective symmetry forbidden modes
\cite{huang_mode_2010}. Figures \ref{fig:position} (a--d) show excitation
efficiency maps for both fundamental polarizations obtained by recording the
integrated emission intensity at the respective positions of the mode detector
as a function of excitation spot position over a range of $1\times1$~$\mu$m. For
parallel (perpendicular) polarization the maps feature a single connected region
roughly centered on the incoupling antenna if the emission is recorded at the
corresponding emission position of the symmetric (antisymmetric) mode. If the
``wrong'' mode detector port emission is recorded, then, despite the seemingly
wrong polarization, the corresponding modes can still be detected if the beam is
displaced perpendicular to the TWTL axis. This results in two disconnected areas
in the excitation efficiency maps which are reproduced by FDTD simulations
[Figs. \ref{fig:position} (e--h)]. We conclude that excitation of pure modes
requires both control of the polarization and nanometer-scale precision for the
positioning of the excitation spot. We further conclude that both degrees of
freedom, polarization and focal spot position, can be used to synthesize linear
combinations of pure modes.

\begin{figure}[htbp]
\begin{center}
\includegraphics[width=\columnwidth]{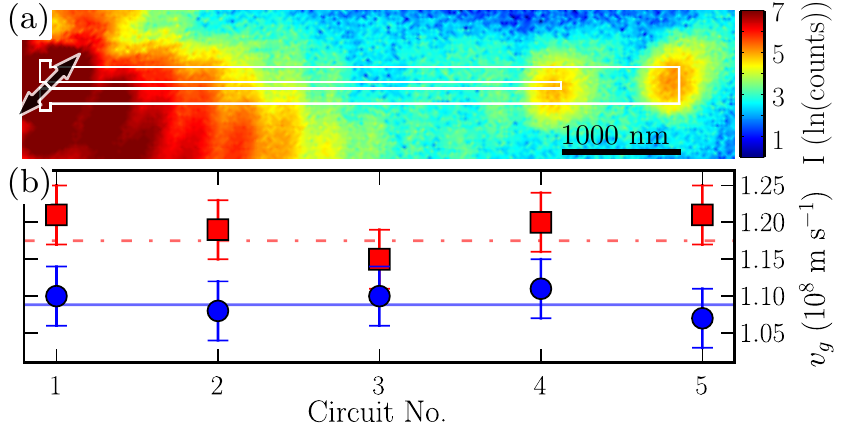}%
\caption{(color online) (a) Emitted intensity scan (log-scale, no attenuation of
	direct scattering) of the circuit using 45$^\circ$ polarized illumination
	($\lambda =$ 800~nm) \cite{rewitz_ultrafast_2012,
	rewitz_spectral-interference_2012}. (b) Group velocities of symmetric (red
	squares, dashed line) and antisymmetric mode contributions (blue circles,
	solid line) of ultrafast plasmon pulses determined for one of the arrays
	used in Fig. \ref{fig:incoupling}. Quantitative agreement between
	experimental data (symbols with error bars) and simulation results
	(horizontal lines) is obtained. Note that the values obtained here differ
	from those in Table \ref{tab:mode-properties} due to different wavelengths.
\label{fig:vg}}
\end{center}
\end{figure}

%
%
As an application we demonstrate the selective determination of the group
velocities of symmetric and antisymmetric contributions of a multimode ultrafast
plasmon pulse. We create such a plasmon pulse using a well-positioned excitation
spot polarized at 45$^\circ$ with respect to the TWTL [Fig. \ref{fig:vg} (a)].
The experiment is performed using ultrashort pulses (800~nm central wavelength,
53~nm FWHM, 80~MHz repetition rate, 2~nW average excitation power) on a
dedicated setup \cite{rewitz_ultrafast_2012, rewitz_spectral-interference_2012}
using one of the TWTL arrays of Fig. \ref{fig:incoupling}. A time-averaging
detector imaging the mode detector records about equal intensities for both
ports. However, since both modes travel at different group velocities, the
symmetric and the antisymmetric pulse contributions actually arrive at their
ports at slightly different instants after correcting for the total propagation
distance. Such minute time delays as well as absolute propagation times can be
measured using spectral-interference microscopy
\cite{rewitz_spectral-interference_2012}. From the determined propagation times
we calculate the respective modal group velocities. The results are displayed in
Fig. \ref{fig:vg} (b) and compared to simulated values. Within the error margins
quantitative agreement between experiment and theory is found and the small
differences in the modal group velocities can be clearly resolved. As expected,
the less-confined symmetric mode is closer to the free-space propagation speed
(about 10\% faster than the antisymmetric mode). No systematic influence of the
antenna length on the pulse propagation time is observed.

We conclude that near-infrared plasmon excitations (cw and ultrafast) in top-down 
fabricated single-crystalline gold TWTL nanocircuits can be prepared in
deterministic eigenmode superpositions by positioning a tightly focused laser
beam with respect to the incoupling antenna attached to the TWTL and by
adjusting its polarization. Modal analysis is performed \emph{in-situ} by means
of a mode detector structure that is read out by a single-shot far-field
measurement providing a direct feedback on the modal composition. 
Finally, we use these abilities to study the propagation of a deterministic 
ultrashort multimodal plasmon pulse by the separate measurement of the group 
velocities of its eigenmode contributions.


\begin{acknowledgments}
The authors gratefully acknowledge Thorsten Feichtner, Monika Emmerling, and
Monika Paw{\l}owska for 3D artwork, for fabricating the beam block,  and
fruitful discussions, respectively as well as the DFG for financial support
within the Priority Program ``Ultrafast Nanooptics'' (SPP 1391).
\end{acknowledgments}

\bibliography{references/references}

\end{document}